%Paper: 9109017
%From: Nir Sochen <sochen@titanic.saclay.cea.fr>
%Date: 11 Sep 91 16:19:25+0200

\input harvmac.tex
\def\J#1#2{J^{#1}_{#2}}
\def\sss{\scriptscriptstyle}
\Title{SPhT/91-117}{Singular Vectors by Fusions in $A_1^{(1)}$}
\centerline{M. Bauer and N. Sochen}
\centerline{\it Service de Physique Th\'eorique CE- Saclay
\footnote{$^1$}{Laboratoire de la Direction des Science de la Mati\`ere
du Commissariat \`a l'Energie Atomique}, F-91191 Gif-sur-Yvette cedex, France}
\vskip3cm
\centerline{ABSTRACT}
Explicit expressions for the singular vectors in the highest weight
representations of $A_1^{(1)}$ are obtained using the fusion formalism
of conformal field theory.
\Date{8/91}
\bigskip
We study in this letter some aspects of the representation theory of
$A_1^{(1)}$.
The subject was studied thoroughly by mathematicians \ref\rK{
See for example:\hfil\break
V. Kac, Infinite
dimensional Lie algebras. Cambridge: Cambridge University Press 1985, and
references therein.}
and physicists \ref\rPH{
Fractional levels appear naturally in 2d quantum gravity:\hfil\break
A.M. Polyakov, Mod. Phys. Lett. A 2 (1987) 893.\semi
V.G. Knizhnik, A.M. Polyakov and A.B. Zamolodchikov, Mod. Phys Lett A 3 (1988)
819.\semi
D. Bernard and G. Felder, Comm. Math. Phys. 127 (1990) 145.}
in the last years. It is known that a Verma module (i.e highest weight
representation) is
reducible for certain values of the highest weight. This indicates the
existence of a singular vector in the module, that generates a submodule.
An irreducible representation is obtained by quotienting the submodules out,
or equivalently by setting the singular vectors to zero.
The later point of view is the most appropriate to physical applications.

Conformal field theory Ward identities permit us to express any correlation
function that contains descendents as a differential operator that acts on the
correlation function with the descendent replaced by its primary \ref\rBPZ{
A. Belavin, A.M Polyakov and A.B. Zamolodchikov, Nucl. Phys. B241 (1983) 333.}.
A singular
vector is a descendent that we equate to zero. We get, therefore, a
differential
equation for any correlation function that includes the corresponding primary
field. The explicit form of this equation is fixed by the form of the singular
vector. Explicit forms for singular vectors in Verma modules for the Virasoro
algebra were obtained recently by Bauer, Di Francesco, Itzykson, and Zuber
\ref\rBDIZ{M. Bauer, Ph.Di Francesco, C. Itzykson and J.-B. Zuber, ``Covariant
Differential Equations and Singular Vectors in Virasoro Representations'',
Saclay Preprint SPht/91-030. to appear in Nucl. Phys. B.\semi
See also\hfil\break
M. Bauer, ``Singular Vectors in Virasoro Verma Modules'', Saclay Preprint
SPht/91-096, to appear in the Proceeding of the Trieste Workshop on String
Theory, April 24-26, 1991, Trieste, Italy.}.
Forms for $A^{(1)}_N$ were
given by Malikov, Feigin and Fuks \ref\rMFF{
F.G. Malikov, B.L. Feigin and D.B Fuks, Funkt.Anal. Prilozhen, 20 No. 2 (1987)
25}.
 The form that M.F.F. found reflects the symmetries of the
$A^1_N$ algebra ,but it is not very suitable to physical application.
Our approach is based on basic facts and methods in conformal field theory,
with which the form of the singular vectors for $A^{(1)}_1$ is derived
efficiently. This approach is in the same spirit of B.D.I.Z. and we have
analogously a mysterious relation between the singular vectors and the
classical w-algebras \rBDIZ.

We present the infinite-dimensional Lie algebra $A^{(1)}_1$ as a current
algebra
. The generators are $\J an\;\; n\in Z\;\; a=+,0,-$ and $K$. The non-vanishing
commutators are
$$\eqalign{
  [\J+n,\J-m]&=2\J0{{n+m}}+nK\delta_{n+m}\cr
  [\J0n,\J\pm m]&=\pm\J\pm{{n+m}}\cr
  [\J0n,\J0m]&={1\over 2}nK\delta_{n+m}\cr}$$
The generators $\J an$ are doubly graded, with respect to $a$ (generated
by $\J00$) and to $n$ (generated by $L_0$).
It is convenient to introduce a combined gradation $d(n,a)=2n+a$.
We are interested in the representations generated by a highest weight vector.
The highest weight vector is defined by
$$\eqalign{\J an |J,t>&=0\;\; d(n,a)\geq 1\cr
            \J00 |J,t>&=J|J,t>\cr
                K|J,t>&=(t-2)|J,t>\;\;  J,t\in {\bf C}\cr}$$
Notice that $J$ is not a half-integer as usual but takes value in the
complex plane, and we work with complex central charge as well. All the
states are reached by applying repeatedly $\J an$ with $d(n,a)\leq -1$ on
the highest weight, and taking linear combinations. A combination
$|R>=\sum A(\{a_i\},\{n_i\})\J{{a_s}}{{n_s}}
\cdots \J{{a_1}}{{n_1}}|J,t>$ is called a descendent.
A descendent $|R>$ that satisfies $\J an |R>=0$ for $d(n,a)\geq 1$ is called
a singular vector. It is easy to see that singular vectors are linear
combinations of homogeneous singular vectors w.r.t the graduation $L_0$
and $\J00$. We denote the homogeneous part as
$$|R>\equiv |J,t,n,m>=\sum_{{a_1+\cdots+a_s= -m}\atop {n_1+\cdots+n_i= n}}
 A(\{a_i\},\{n_i\})\J{{a_s}}{{n_s}}\cdots \J{{a_1}}{{n_1}}|J,t>$$

$\underline{\rm Theorem}$: (Kac-Kazhdan \ref\rKK{
V.G. Kac and D.A. Kazhdan, Adv. Math 34 (1979) 97.})

The highest weight $|J,t\neq0>$ generates a reducible module iff
$J$ takes the value $J_{r,s,\pm}$ defined by
\eqn\eKK{2J_{r,s,\pm}+1=\pm(r-(s-1)t)\;\;r,s\in{\bf N}}
The singular vector is in $|J_{r,s,\pm},t\neq0,r(s-1),\pm r>$.

$\underline{\rm Theorem}$ (Malikov,Feigin and Fuks \rMFF\ )

The form of the singular vector in $|J_{r,s,+},t\neq0>$ is
\eqn\eS{(\J-0)^{r+(s-1)t}(\J+{-1})^{r+(s-2)t}\ldots(\J+{-1})^{r-(s-2)t}(\J-0)^
{r-(s-1)t}|J_{r,s,+},t\neq0>}
and $\J-0\leftrightarrow\J+{-1}$ for the sign $(-)$ in \eKK .
This is not to be interpreted as multiplication of operators raised
to complex powers, but in the following way: for $t\in {\bf N}$ \eS\ is well
defined, and by ordering the generators and applying the Poincar\'e Birkhoff
Witt theorem, this vector can be expressed as
\eqn\eP{
\sum_{p,q=0}^{\infty}P_{p,q}(\{\J a{-n}\},t)
(\J+{-1})^{r(s-1)-p}(\J-0)^{rs-q}|J_{r,s,+},t>}
where $p,q$ are non negative integers, $P_{p,q}$ are polynomials in the
$\J an$ with $d(n,a)\leq-2$. The heart of the proof is to show that $P_{p,q}$
are polynomials in $t$, and this allows to analytically continue to the
whole complex plane.  Actually calculating the r.h.s of \eP , by applying
successively the analytical continued commutation relations turns out to be
 tedious. It is the aim of this letter to give a short-cut for it.

Let us consider for the moment only the + sign in \eKK\ then $J=j-j^\prime t$
where $2j+1,2j^\prime+1 \in {\bf N}$. We denote the corresponding highest
weight representation as $(j,j^\prime)$. The singular vector in the class
of representation of type $(0,j)$ admits a matrix form, the $A_1^{(1)}$ analog
of the formulae of Benoit and Saint-aubin \ref\rBSA{
L. Benoit and Y. Saint-Aubin, Phys. Lett. 215B (1988) 517.}
and of B.D.I.Z in the case of Virasoro.

Let $n=2j+1$. We define the $2n-1$ dimensional vectors:
$$\eqalign{\vec f&= (f_{J+n-1},g_{J+n-1},f_{J+n-2},\cdots,g_{J+1},f_J)^t\cr
\vec F &=(-g_{J+n},0,\cdots,0)^t\cr}$$
where $f_J\equiv |J,t>$ and the other components are defined by
solving the triangular system \eqn\eMF{\vec F= \left(J^- +\sum_{s=0}^\infty
R(s+1)t^sK^s\right)\vec f}
where
$$\eqalign{
(J^-)_{i,j}&=\delta_{j,i+1}\;\;\; i,j=1,\ldots,2n-1\cr
R(s+1)_{i,j} &=(\J0{-{{s+1}\over 2}}\delta^{(2)}_{s,1}+
(\J+{-{{s+2}\over 2}}\delta^{(2)}_{i,0}+\J-{-{s\over2}}\delta^{(2)}_{i,1})
\delta^{(2)}_{s,0})(-1^{[{s\over2}]})\delta_{i,j}\cr
(K)_{i,j} &= ((n-{{i+1}\over2})\delta^{(2)}_{i,1}+{i\over2}\delta^{(2)}_{i,0})
\delta_{j,i+1}\cr}$$
where $[x]$ is the integer part of x, and
$$\delta^{(2)}_{i,j}=\cases{1 & $i=j$ mod 2\cr 0& otherwise\cr}$$
In fact $K^{2n-1}\equiv0$, so the sum in \eMF\ is finite.
One can show that the components of $\vec f,\vec F$ satisfy
$$\eqalign{\J+0 f_{J+r}&=0\hskip 3cm\J-1 f_{J+r}=-rtg_{J+r}\cr
  \J+0 g_{J+r}&=(n-r)tf_{J+r-1}\hskip 1cm \J-1 g_{J+r}=0\cr}$$
from which is clear that $g_{J+n}$ is a singular vector. Take for example
the representation $(0,{1\over2})$ (that is $J=-{1\over2}t$), then \eMF
reads
$$\pmatrix{-g_{J+2}\cr 0\cr 0\cr}=
\pmatrix{\J-0 & t\J0{-1} & -t^2\J-{-1}\cr
           1  &\J+{-1}   & t\J0{-1}   \cr
           0  &  1       & \J-0       \cr}
\pmatrix{f_{J+1} \cr g_{J+1} \cr f_J \cr}$$
which gives the singular vector
$$g_{J+2}= (\J-0\J+{-1}\J-0 -t\J-0\J0{-1} -t\J0{-1}\J-0 -t^2\J-{-1})f_J$$

Notice that in the ``classical limit'', and after ``gauging''  in the spirit of
Hamiltonian reduction
$$\eqalign{
\J-0&\to d\cr
\J0{-n}&\to 0\cr
\J+{-n}&\to 0\;\;\; n\neq -1\cr
\J+{-1}&\to 1\cr
(-1)^{[{{n+1}\over2}]}t^{2n}\J-{-n}&\to w_{n+1}\cr}$$
this matrix, that maps $f_J$ to $g_{J+n}$, becomes the covariant differential
operator of Drinfeld-Sokolov \ref\rDS{
V.G. Drinfeld and V.V. Sokolov, Journ. Sov. Math. 30 (1985) 1975.}
that maps $(1-n)/2$ forms to $(1+n)/2$ forms.

We show now how the matrix form appears naturally from fusions of primary
fields in conformal field theory. Our main result will be a recursive
formula for the components of $\vec f$ from which the matrix can be recovered.
The advantage of this approach is that it gives us also the form of the
singular vector in a general $(j,j^\prime)$ reducible representation. It
gives us important information about fusion rules as well.

We introduce a chiral primary field $\phi^J_m(z)$ w.r.t the Virasoro algebra
as well as w.r.t $A_1^{(1)}$. It transforms as a vector under the horizontal
algebra (the zero modes algebra):
$$[\J a0,\phi^J_m(z)]= R^a_{mn}\phi^J_n(z)$$
Following Zamolodchikov and Fateev \ref\rFZ{
A.B. Zamolodchikov and V.A. Fateev, Sov. J. Nucl. Phys. 43 (1986) 657.}
we introduce an auxiliary parameter in order
to have
$$[\J a0,\phi^J(z,x)]= R^a(x)\phi^J(z,x)$$
where $R^a(x)$ is a differential operator.

The correspondence fields-states is given by
$\lim_{z\to 0}\lim_{x\to 0}\phi^J(z,x)|\Omega,t>=|J,t>$. Here $|\Omega,t>$ is
the
vacuum which is characterized as a highest weight state that is
annihilated by the whole horizontal algebra.

The Virasoro algebra and $A_1^{(1)}$ are related by the Sugawara construction
$$L_n=\sum(:J^0_{n-m}J^0_m+{1\over2}J^+_{n-m}J^-_m+{1\over2}J^-_{n-m}J^+_m:)$$
In this formulation $L_{-1}$ and $J^-_0$ generate translations in z and in x
respectively. Thus, we can write
$$\phi^J(z,x)=e^{xJ^-_0+zL_{-1}}\phi^J(0,0)e^{-xJ^-_0-zL_{-1}}$$
It follows that
\eqna\eCR
$$\eqalignno{
[J^-_n,\phi^J(z,x)]&=z^n{d\over dx}\phi^J(z,x)&\eCR a\cr
[J^0_n,\phi^J(z,x)]&=z^n(-x{d\over dx}+J)\phi^J(z,x)&\eCR b\cr
[J^+_n,\phi^J(z,x)]&=z^n(-x^2{d\over dx}+2xJ)\phi^J(z,x)&\eCR c\cr}$$
Let us look now at the short distance operator product expansion for these
chiral primary fields. For this aim it is more convenient to write
$$\phi^J(z,x)=z^{-h+L_0}x^{J-J^0_0}\phi^J(1,1)x^{J^0_0}z^{-L_0}$$
which is a consequence of \eCR{b}. Imagine that we are interested only in the
$J$ sector in the fusion of $J_0$ and $J_1$ then
\eqn\eF{
\eqalign{\phi^{J_0}(z,x)|J_1,t>&=\phi^{J_0}(z,x)\phi^{J_1}(0,0)|\Omega,t>\cr
&=z^{-h_0-h_1+L_0}x^{J_0+J_1-\J00}\phi^{J_0}(1,1)\phi^{J_1}(0,0)|\Omega,t>\cr
&\buildrel{in\ sector\ J}\over{\hbox to 30pt{\rightarrowfill}}
 \sum_{n=0}^{\infty}\sum_{m=-n}^{\infty}
z^{h-h_0-h_1+n}x^{-J+J_0+J_1+m}
\psi^{J}_{n,m}(0,0)|\Omega,t>\cr}}
$\psi^J_{n,m}$ are fixed by the requirement that the two sides of \eF\
transform in the same manner under the Vir and $A_1^{(1)}$ algebras. Thus, for
example
$$\eqalign{\phi^{J_0}(z,x)L_0\phi^{J_1}(0,0)|\Omega,t>&=
h_1\phi^{J_0}(z,x)\phi^{J_1}(0,0)|\Omega,t>\cr
&=(L_0-{d\over{dz}})\phi^{J_0}(z,x)\phi^{J_1}(0,0)|\Omega,t>}$$
Plugging \eF, we obtain
$$L_0\psi^J_{n,m}=(h+n)\psi^J_{n,m}$$
and in the same way
$$\J00\psi^J_{n,m}=(J-m)\psi^J_{n,m}$$
This fixes $\psi^J_{n,m}$ to be in the homogeneous subspace of conformal
degree n and charge -m, as expected from \eF\ . For the generators
$\J an\;\;d(n,a)\geq 1$ there are only two independent equations
\eqn\eD{\eqalign{
\J+0\psi^J_{n,m}&=(J-J_0+J_1-m+1)\psi^J_{n,m-1}\cr
\J-1\psi^J_{n,m}&=(-J+J_0+J_1+m+1)\psi^J_{n-1,m+1}\cr}}
These descent equations determine $\psi^J_{n,m}$ as long as the kernel of
$\J+0$,$\J-1$ is trivial. The existence of a singular vector in $|J,t,n_0,m_0>$
means that the kernel of $\J+0,\J-1$ is non trivial, and as a consequence
$J_0$, $J_1$ and $J$ must verify a relation in order for \eD\ to have a
solution. This relation is the {\it fusion rule}. Take the
vacuum sector for example, $\psi^0_{\sss0,0}\equiv|\Omega,t>$ and
$\psi^0_{\sss0,1}=A_{\sss0,1}\J-0\psi^0_{\sss0,0}$ is a singular vector.
The non-trivial descent equation gives
$0=\J+0\psi^0_{\sss0,1}=(J_1-J_0)\psi^0_{\sss0,0}$ and we see
that only the fusion of a primary field with itself contains the vacuum sector.

A solution to the descent equations is obtained using the Knizhnik-
Zamolodchikov equation \ref\rKZ{
V.G. Knizhnik and A.B. Zamolodchikov, Nucl. Phys. B247 (1984) 83.}
combined with the fusion procedure. We write
\eqn\eKZA{
(tL_{-1}-(\J+{-1}\J-0+2J_1\J0{-1}))\phi^{J_1}(0,0)|\Omega,t>=0}
multiply on the left by $\phi^{J_0}(z,x)$ to get
\eqn\eKZ{0=
\phi^{J_0}(z,x)(tL_{-1}-(\J+{-1}\J-0+2J_1\J0{-1}))\phi^{J_1}(0,0)|\Omega,t>}
We commute the $A_1^{(1)}$ and Vir operators to the left using \eCR{} ,
plugging \eF, and after some manipulations
we obtain
\eqn\eRF{\eqalign{(nt+m(2J+1-m))\psi^J_{n,m}=
(-J+J_0+J_1+m+1)&\sum_{k+l=n\atop k\geq1}\J+{-k}\psi^J_{l,m+1}\cr
+2(J-J_0-m)&\sum_{k+l=n\atop k\geq1}\J0{-k}\psi^J_{l,m}\cr
+(J-J_0+J_1-m+1)&\sum_{k+l=n\atop k\geq0}\J-{-k}\psi^J_{l,m-1}\cr}}
If $nt+m(2J+1-m)$ does not vanish for any pair of integers $n\geq0\;\;
m\geq -n$ then the $\psi^J_{n,m}$ are uniquely determined starting from
$\psi^J_{\sss 0,0}$, and can be shown recursively to satisfy the descent
equations.
If $J$ is such that $n_0t+m_0(2J+1-m_0)=0\;\;\;n_0\geq0\;\;m_0\geq -n_0$ the
same is true provided $n<n_0$ or $m+n<m_0+n_0$, and for $(n,m)=(n_0,m_0)$, the
right hand side of \eRF\ is annihilated by $\J+0$ and $\J-1$. In this case
there are two possibilities. If $m_0$ does not divide $n_0$ then both sides of
\eRF vanish.  If $m_0$ divides $n_0$ then the r.h.s is not trivial in general,
and it is in the kernel of $\J+0$ and $\J-1$, that is, it is the wanted
singular vector.  The descendents $\psi^J_{n,m}$
$0<n<n_0$, $ -n<m<2n_0-n$ and $-n_0<m<m_0$ for $n=n_0$ can be arranged as
a column vector. Equation \eRF\ then can be written in a matrix form,
analogous to \eMF. Details and proofs will be given elsewhere. Quantum
Hamiltonian reduction \ref\rQHR{
A.A Belavin, Advanced Studies in Pure Mathematics 19 (1989) 117. \semi
M. Bershadsky and H.Ooguri, Comm. Math. Phys. 126 (1989) 49. \semi
For approach akin to ours, see\hfil\break
P. Furlan, A.Ch. Ganchev, R. Paunov and V.B. Petkova, ``Reduction of the
Rational Spin {\it sl(2,C)} WZNW conformal Theory'', Preprint
SISSA-67/91/FM, KA-THEP-1991-3.}
between the physical spaces of Vir and $A_1^{(1)}$, and
the relation between the $A_1^{(1)}$ singular vectors and classical w-algebra
are under investigation.

\listrefs
\bye